\documentclass[12pt,preprint]{aastex}

\newcommand\Mjup{\mbox{$M_{\rm Jup}$}}

\newcommand\Msun{\mbox{$M_\sun$}}

\newcommand\masyr{{\rm mas~yr\mbox{$^{-1}$}}}
\newcommand\maspix{mas~pix\mbox{$^{-1}$}}

\begin{document}

\shortauthors{Metchev et al.}
\shorttitle{Pre-Discovery Image of HR 8799}

\title{Pre-Discovery 2007 Image of the \object{HR 8799} Planetary System}

\author{Stanimir Metchev}
\affil{Department of Physics and Astronomy, State University of New York, Stony Brook, New York 11794--3800}
\email{metchev@astro.sunysb.edu}
\author{Christian Marois}
\affil{National Research Council Canada, Herzberg Institute of Astrophysics, 5071 West Saanich Road, Victoria, BC V9E 2E7, Canada}
\and
\author{B.~Zuckerman}
\affil{Department of Physics \& Astronomy, University of California, Los Angeles, California, 90095--1562}

\begin{abstract}
We present a pre-discovery $H$-band image of the \object{HR 8799} planetary system
that reveals all three planets in August 2007.  The data were obtained with the Keck adaptive optics system, using angular differential imaging and a coronagraph.  We confirm the physical association of all three planets, including HR~8799d, which had only been detected in 2008 images taken two months apart, and whose association with HR~8799 was least secure until now.  We 
confirm that the planets are 2--3~mag fainter than field brown dwarfs of comparable near-infrared colors.
We note that similar under-luminosity is characteristic of young substellar objects at the L/T spectral type transition, and is likely due to enhanced dust content and non-equilibrium CO/CH$_4$ chemistry in their atmospheres. 
Finally, we place an upper limit of $\gtrsim$18~mag per square arc second on the $>$120~AU $H$-band dust-scattered light from the HR~8799 debris disk.  The upper limit on the integrated scattered light flux is $10^{-4}$ times the photospheric level, 24 times fainter than the debris ring around HR~4796A.
\end{abstract}


\section{INTRODUCTION}

The three-planet system around the A5V star HR~8799 \citep[][henceforth, M08]{marois_etal08b} is the first directly imaged extrasolar multi-planet system.  Along with a reported extrasolar planet around Fomalhaut \citep[A4V;][]{kalas_etal08}, HR~8799b, c, and d are also the first bona-fide extrasolar planets directly imaged around stars other than the Sun.
The HR~8799 and Fomalhaut planetary systems share two important similarities.  First, in both cases the planets are widely separated from their host stars, at projected separations between 24--119~AU.  And second, both hosts are A stars surrounded by cold debris disks \citep{aumann85, sadakane_nishida86}, circumscribing the planetary systems.  

The debris disk around HR~8799 is one of the most massive detected by {\it IRAS} and is a factor of several brighter \citep[$L_{\rm IR}/L_\ast=2.3\times10^{-4}$;][]{moor_etal06, rhee_etal07} than that around Fomalhaut \citep[$L_{\rm IR}/L_\ast=8\times10^{-5}$;][and references therein]{rhee_etal07}.  While Fomalhaut's disk has been spatially resolved at wavelengths ranging from the visible \citep{kalas_etal05} to the submillimeter \citep{holland_etal98, marsh_etal05}, the HR~8799 debris disk remained unresolved until recently \citep[][henceforth, S09]{su_etal09}.

Prior to the discovery of the planets around HR~8799 by M08, we targeted the star with the Keck adaptive optics (AO) system \citep{wizinowich_etal00} to detect the debris disk in scattered light.  The strength of its IR excess and a detection in the submillimeter \citep[$\approx$10~mJy at 850~$\micron$;][]{williams_andrews06} indicated a substantial optical depth, which given suitable disk viewing geometry could result in a detection of dust-scattered light.  The data presented here did not produce the sought-after scattered light detection, but have revealed all three known planets around HR~8799.  In particular, we report the earliest-epoch image of the closest-in planet, HR~8799d. 
Prior epoch detections of HR~8799b and/or HR~8799c have been published in M08, \citet{fukagawa_etal09}, and \citet{lafreniere_etal09}. 

\section{OBSERVATIONS \label{sec_observations}}

We observed HR~8799 ($H=5.28$~mag; 39.4~pc) with the Keck~II AO system on 2 August 2007.  Atmospheric seeing was variable, ranging from $0\farcs6-1\farcs4$ in the visible.  Cloud cover was pervasive and ranged from contributing $\sim$1~mag in visual extinction to at times completely preventing telescope operations.  Observations were taken at $H$ band in angular differential imaging \citep[ADI;][]{marois_etal06} mode, by fixing the telescope pupil rotator to optimize contrast and the detectability of non-face-on disks.  
We used the NIRC2 near-IR camera in its 10~\maspix\ (``narrow'') scale for optimal spatial sampling with the ADI technique. 
We targeted HR~8799 for a total of five hours around transit with frequent weather interruptions.  
We obtained 114 one-minute exposures, 
of which 73 (i.e., 1.22~hrs total) were
adequate for high-contrast science.  All were taken with the star covered by the 1$\farcs$0-diameter semi-transparent NIRC2 coronagraph, and none were saturated. 
The LARGEHEX hexagonal pupil plane stop within NIRC2 was used to mask the edges of the primary mirror.
Several unocculted AO-corrected images (5~ms, 1000 coadds per integration) were acquired during periods of stable seeing to calibrate the coronagraph's transmission.
At flux levels $\leq$8,500 cts pix$^{-1}$ coadd$^{-1}$ HR~8799 was in the linear response regime of the camera in all unocculted exposures.

Sets of five 60~s coronagraphic exposures of a point-spread function (PSF) reference star, HR~8788 (F6V, $H=5.13$~mag),
were obtained at three different times during the night for additional PSF calibration.  Images of the dome-illuminated interior with the 1$\farcs$0 coronagraph in  place were used for flat-field calibration.  One-minute dark frames were taken at the beginning of the night.

\section{DATA REDUCTION \label{sec_data_reduction}}

\begin{figure}
\plotone{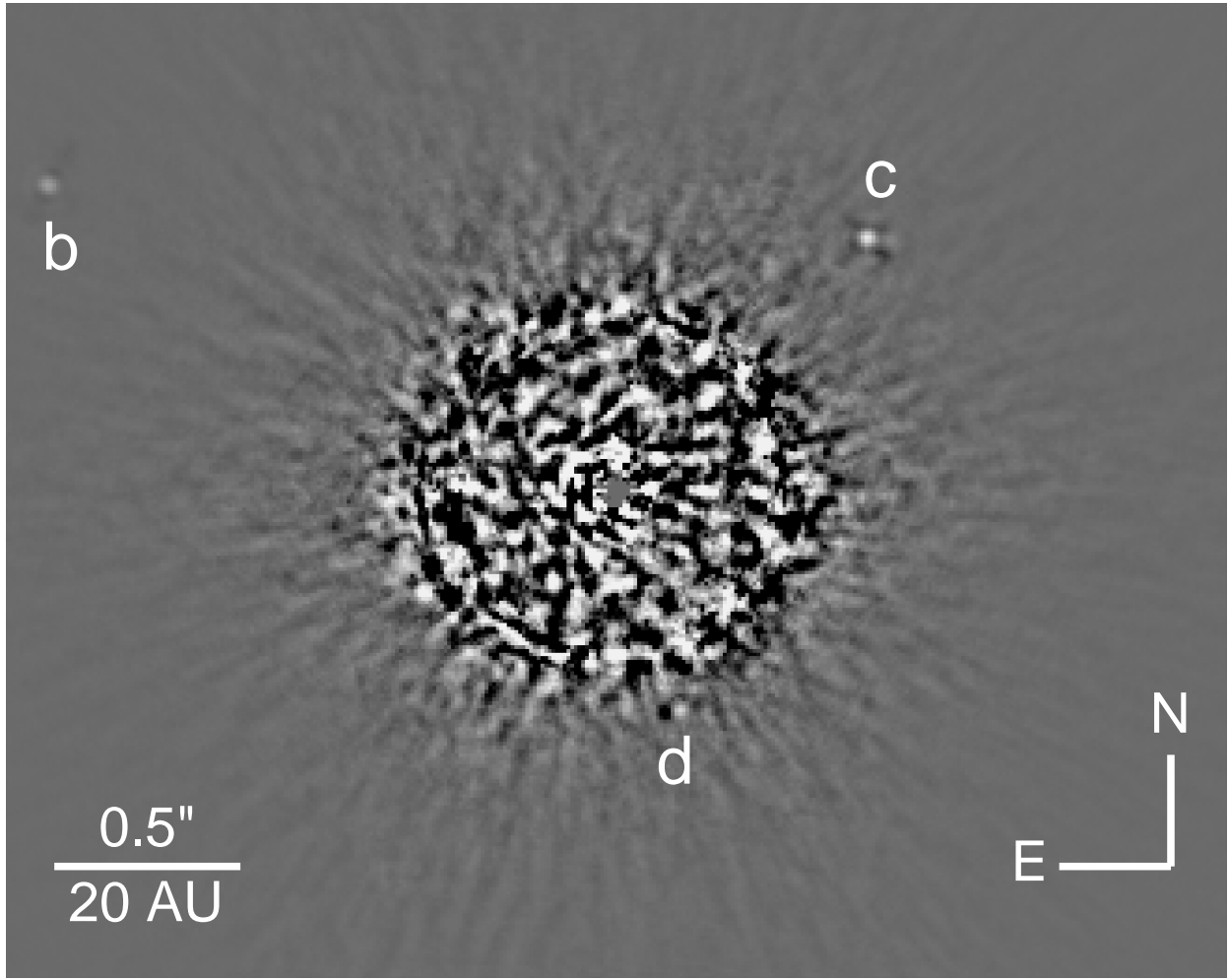}
\figcaption{Keck AO $H$-band image of the HR~8799bcd planetary system from 2 August 2007.  The identification of HR~8799d in this image is the earliest by almost a year.
\label{fig_hr8799_image}}
\end{figure}

The ADI sequences of exposures were dark-subtracted, flat-fielded, bad-pixel corrected, and reduced using custom-made ADI and LOCI \citep[``locally optimized combination of images'';][]{lafreniere_etal07} routines. The images were then spatially cross-registered and averaged to produce the final image shown in Figure~\ref{fig_hr8799_image}.  Precise frame registration 
was possible because of the partially transmissive property of the NIRC2 coronagraph, which allowed alignment on HR~8799 itself.  
The final image is comparable in contrast to those published in M08, and all three known planets are detected. 

ADI observations optimize contrast for non-azimuthally symmetric circumstellar structure, such as point source companions or a non-face-on disk.  However, because the HR~8799bcd planetary system is seen near pole-on, and given the likelihood of planar alignment between the planetary orbits and the  
debris disk, we also independently reduced our data using the PSF reference star exposures. To mitigate the effect of variable AO performance 
we applied a radially-dependent scaling to the reference PSF, enhancing a procedure used in \citet{metchev_etal05}.  
More precisely, after registering the science and the reference-star exposures, we fit a linear relation $f_{ij}(r)$ to the ratio $q_{ij}(r)=s_i(r)/p_j(r)$ of the radial profiles of each science exposure $i$ with the nearest PSF exposure $j$.  We then scaled the PSF by $f_{ij}(r)$, and subtracted the thus re-normalized PSF from the science exposures before averaging. 
The linear fit to $q_{ij}(r)$ was obtained at $\delta r=80$~mas steps over overlapping 3$\arcsec$-wide search annuli.  The PSF subtraction was thus 
aimed at revealing high order (quadratic or more) radial structure with widths up to $\sim$3$\arcsec$.   The PSF-subtracted exposures were rotated, aligned, and then averaged and 3$\sigma$-clipped to produce the final image.

Even with the use of a separate PSF reference star, we did not detect scattered light from the debris disk around HR~8799.  
This type of data reduction does show HR~8799b and c, although it does not reveal the innermost planet d.

\section{ANALYSIS AND DISCUSSION}


\subsection{Astrometry of the HR 8799 Planetary System}

\begin{figure}
\plotone{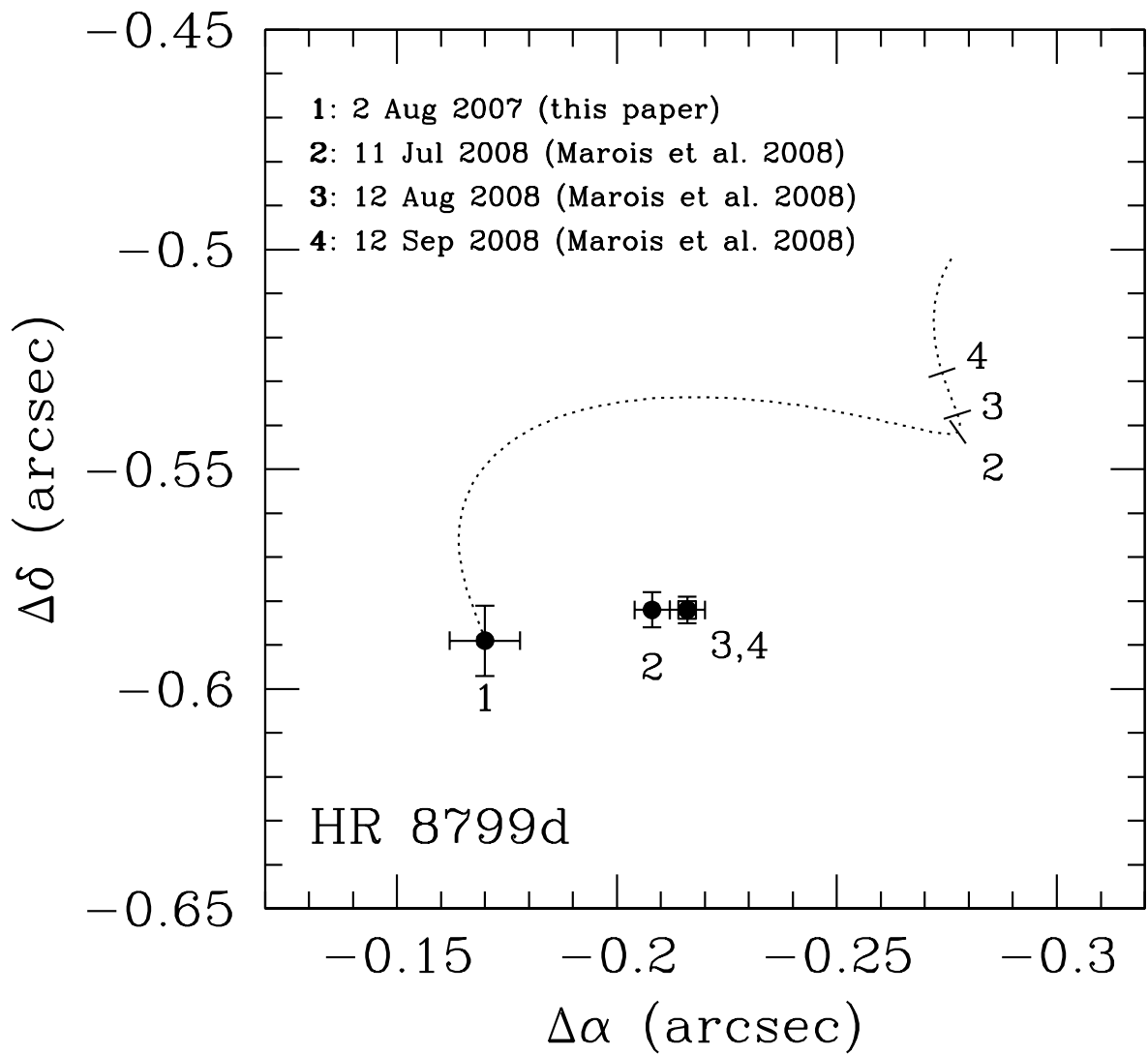}
\figcaption{Orbital motion of HR~8799d between 2 August 2007 (position {\bf 1}) and 12 September 2008 (position {\bf 4}) relative to the position of HR~8799.  The dotted line with perpendicular dashes at the corresponding observational epochs traces the expected change in position of HR~8799d due to the proper and parallactic motion of the primary if the planet were an unrelated background object.
\label{fig_astrometry}}
\end{figure}

\begin{deluxetable}{lrrll}
\tabletypesize{\scriptsize}
\tablewidth{0pt}
\tablecaption{Astrometry and $H$-band Photometry for HR~8799bcd from 2 August 2007 \label{tab_astro_phot}}
\tablehead{ & \colhead{$\Delta\alpha$} &
\colhead{$\Delta\delta$} & \colhead{$H$} &
\colhead{$H_{\rm M08}$\tablenotemark{\dag}}\\
\colhead{Planet} & \colhead{(arcsec)} &
\colhead{(arcsec)} & \colhead{(mag)} & \colhead{(mag)}}
\startdata
b & $1.522\pm0.003$ & $0.815\pm0.003$ & $18.06\pm0.13$ & $17.85\pm0.17$\\
c & $-0.672\pm0.005$ & $0.674\pm0.005$ & $16.95\pm0.16$ & $16.91\pm0.17$\\
d & $-0.170\pm0.008$ & $-0.589\pm0.008$ & $17.25\pm0.20$\tablenotemark{\ddag} & $16.84\pm0.22$
\enddata
\tablenotetext{\dag}{Apparent $H$-band magnitudes from M08.}
\tablenotetext{\ddag}{The magnitude of HR~8799d may carry additional systematic uncertainty because of its proximity to the coronagraphic mask.}
\tablecomments{
Additional prior-epoch $H$-band photometry for HR~8799b: F160W $=18.54\pm0.12$~mag \citep[from the {\it HST} in 1998;][]{lafreniere_etal09}; $H=18.07\pm0.25$~mag \citep[in 2002;][]{fukagawa_etal09}.}
\end{deluxetable}

We measured the positions of the three planets by fitting two-dimenational Gaussians to their profiles in the reduced ADI image.  Their relative astrometry with respect to HR~8799 is listed in Table~\ref{tab_astro_phot}.  We adopted a pixel scale of $9.963\pm0.005$~\maspix\ for the NIRC2 narrow camera, a $0.13\pm0.02\degr$ offset from North for the camera columns, and corrected for the known camera distortion \citep{ghez_etal08}.  
Astrometric errors were derived by calculating the registration bias from the residual image noise found near the location of each planet.

The measured positions confirm the associations of all three planets with HR~8799, and are consistent with counter-clockwise orbital motion along nearly face-on orbits.
Due to a merely two-month time span between the discovery and confirmation epochs, HR~8799d had the lowest significance (6$\sigma$) of physical association with the host star in M08.  It is now confirmed as a bound object at the $\sim$13$\sigma$ level from the ensemble of astrometric measurements from four independent epochs (Fig.~\ref{fig_astrometry}).
HR8799d was also the planet with the least accurate orbital motion estimate in M08 ($42\pm27$~\masyr).
The longer time-line afforded by the present observations indicates an orbital motion of $41\pm9$~\masyr ($1.6\pm0.4$~AU~yr$^{-1}$).  For a semi-major axis of 24.3~AU and a primary mass of 1.5~\Msun, the orbital period is 98 years and the expected motion along a circular face-on orbit is 1.57~AU~yr$^{-1}$. 

\subsection{Photometry of HR~8799bcd}

Photometry on the planets was performed on the reduced ADI image as described in M08.  We calibrated the planet fluxes relative to that of HR~8799 \citep[$H=5.280\pm0.018$~mag;][]{skrutskie_etal06} seen through the 1$\farcs$0-diameter NIRC2 coronagraph, for which we measured an attenuation of $\Delta H=7.78\pm0.10$~mag
from comparisons with the short unocculted exposures: in agreement with earlier calibrations performed in \citet{metchev_hillenbrand09}.
The ADI/LOCI algorithm and field rotation during exposures were also calibrated by introducing artificial field-rotated planets using the unocculted PSFs. The apparent $H$-band magnitudes for HR~8799bcd are shown alongside those from M08 in Table~\ref{tab_astro_phot}.

Our photometry for HR~8799b and c is in good agreement with that published in M08.  The slightly better precision of our measurements is due to the simultaneous calibration with respect to HR~8799 itself.  M08 did not use a coronagraph for their $H$-band imaging and calibrated their photometry from non-simultaneous data: from short sequences of shallow unsaturated exposures taken between the long sequences of saturated deep exposures.  Our setup avoids non-simultaneous calibration of the data, which can result in difficult to quantify seeing/sky throughput variations. 
HR~8799d is 0.4~mag (1.4$\sigma$) fainter than reported in M08.  While the difference could be random or could indicate real variability in HR~8799d, it could also be caused by a possible decrease in throughput of the optical system in close proximity ($\sim$$0\farcs1=10$~pix $\approx 3$~FWHM) to the coronagraph.
$H$-band photometry of HR~8799b obtained at earlier epochs
(see Table~\ref{tab_astro_phot}) is consistent with the measurements presented both here and in M08.  Hence, 
we do not see evidence for significant variability in any of the planets.


Compared to (mid-T) brown dwarfs of similar intrinsic brightness ($M_H\sim15$~mag), HR~8799b, c, and d appear unusually redder, by $\gtrsim1$~mag in $H-K$, and are matched only by the $\sim$8~Myr-old 2MASS~1207334--393254B \citep[$H-K\sim1.2$~mag;][]{chauvin_etal04}.  Conversely, all four of these planetary-mass objects are 2--3~mag fainter at $H$ band than comparably red $>$1~Gyr-old mid- to late-L field dwarfs.

Similar under-luminosity with respect to the field substellar population, although not as extreme (by $\sim1$~mag at $H$), is already known in $\sim$300~Myr-old brown dwarfs at the L/T spectral type transition, 
where it has been interpreted as evidence of unusually low effective temperatures \citep[$\leq$1200~K;][]{metchev_hillenbrand06, luhman_etal07b}. The warmer than expected spectroscopic appearance of such young cool substellar objects may be linked to unusually high dust content in their low-gravity atmospheres (M08; \citealt{lafreniere_etal09}).  In addition, the fact that the phenomenon is observed in all young L/T-transition dwarfs and only in them (and, to a lesser extent, in a moderately young mid-T brown dwarf binary; \citealt{liu_etal08}) hints at an augmented role of CO/CH$_4$ non-equillibrium chemistry \citep{saumon_etal06} at low substellar surface gravities.  
This hypothesis is further supported by the recent non-detection of the HR~8799 planets at $M$-band by \citet{hinz_etal09}, suggesting unusual strength of the CO fundamental bandhead at 4.7$\micron$.

\subsection{Upper Limits on Outer Planets \label{sec_planet_limits}}

\begin{figure}
\plotone{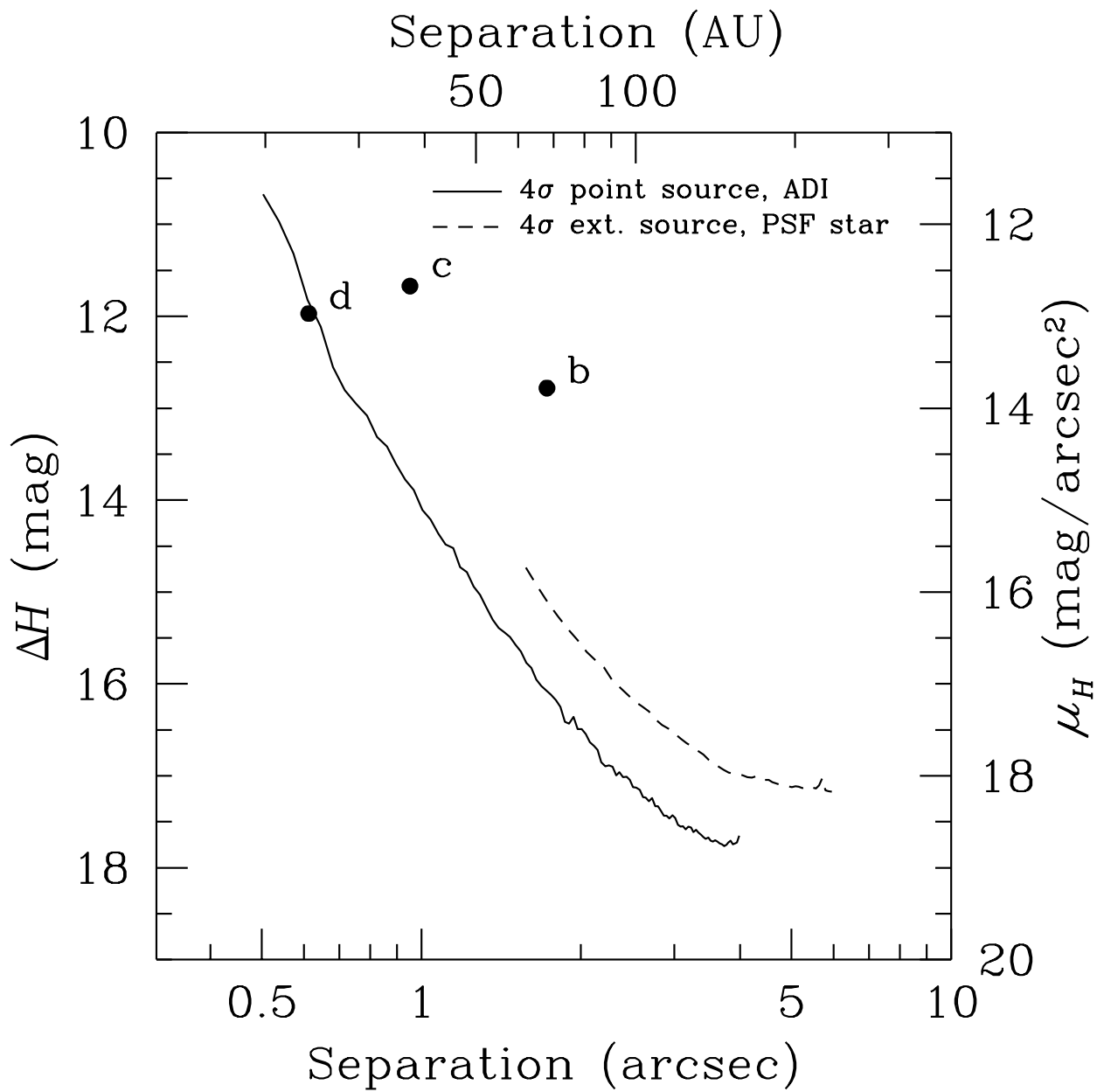}
\figcaption{Detection limits for point sources (in magnitude contrast $\Delta H$ with respect to the primary; solid line) and extended emission (in magnitudes $\mu_{H}$ per square arc second; dashed line) around HR~8799 ($H=5.28$~mag).
The three detected planets are marked.  The point source limits are from the ADI image reduction approach.  The extended source limits are obtained with the use of HR~8788 for PSF calibration.
\label{fig_detlims}}
\end{figure}

The $10\farcs2\times10\farcs2$ NIRC2 field of view (FOV) and a $\sim$1$\arcsec$ off-center positioning of HR~8799
allows us to search for planets out to $\sim$4$\arcsec$ from the star.  We do so by considering the pixel flux scatter in our ADI-reduced final image as a function of radial separation from the star (Fig.~\ref{fig_detlims}).  \citet{marois_etal08} have shown that the residuals obtained from averaging many PSF-subtracted exposures taken with the ADI technique behave in a Gaussian fashion, and so the detectability of faint point sources in the contrast-limited regime can be rigorously determined from the scatter of the pixel fluxes.  To account for the spatial correlation among neighboring pixels on the scale of a PSF, we convolve the ADI-reduced image with the PSF (FWHM~= 3.4~pix~= 34~mas at $H$ band after using a high-pass filter in the data reduction).

We adopt a 4$\sigma$ detection limit for point sources as a level of significance that closely matches the ability to visually identify faint companions \citep{metchev_hillenbrand09}.  Incidentally, we note that 4$\sigma$ is the approximate level at which we detect HR~8799d (Fig.~\ref{fig_detlims}).
We do not find any $H<22$~mag planets beyond the 68~AU projected separation of HR~8799b, complementing results from \citet{close_males09}, who find no $H<22$~mag planets between 5--15$\arcsec$ (200--600~AU). Therefore, for an adopted 60~Myr age of HR~8799, we can exclude additional $>$3~\Mjup\ planets \citep{burrows_etal01, baraffe_etal03} out to at least 600~AU.

\subsection{Debris Disk \label{sec_discussion_disk}}

Given the expectation for a face-on debris disk around HR~8799, we only consider the PSF subtraction performed with the use of a separate PSF star, as that is better suited for revealing 
uniform extended emission.  We disregard the inner region of the planetary system, and focus only on angular separations $>$$1\farcs5$ ($>$60~AU), where the ratios $q_{ij}(r)$ of the HR~8799 to PSF radial profiles
behave in a nearly linear fashion out to the edge of the FOV.  S09 have recently resolved extended 24~$\micron$ and 70~$\micron$ emission from dust orbiting HR~8799 at such separations, which they model as arising from a cool planetesimal belt beyond 90~AU.

Analogous to the procedure adopted for point source detection (\S~\ref{sec_planet_limits}), we convolve our reduced image with the PSF and impose a 4$\sigma$ detection threshold.  We add in quadrature to this (random) source of error
an estimate of the systematic uncertainty $\sigma_{\rm sys}(r)$ in fitting the large-scale wings of the PSF under varying seeing conditions, i.e., in determining $f_{ij}(r)$.  $\sigma_{\rm sys}(r)$ is obtained as the standard deviation of the mean of $f_{ij}(r) p_j(r)$, and 
we find that it contributes approximately 50\% of the error budget. 

We do not detect any $H$-band scattered light from the HR~8799 debris disk down to a limit of $\mu_H\approx$18~mag (65~$\mu$Jy) per square arc second at 3--6$\arcsec$ (120--240~AU; Fig.~\ref{fig_detlims}).  In particular, we do not detect near-IR scattered light from the resolved far-IR disk seen in S09, although it is possible that our 3$\arcsec$-wide search annulus may be over-resolving the observed extended emission.  Given our $10\farcs2\times10\farcs2$ FOV, a wider search annulus was not practical.

It is instructive to compare the upper limit on the surface brightness of the HR~8799 debris disk to other debris disks seen in scattered light, and in particular to the ring around HR~4796A, resolved at 1.6~\micron\ (F160W) with the {\sl HST} by \citet{schneider_etal99}.  For simplicity of comparison, we assume that the HR~8799 debris disk also has a ring-like morphology, with an aspect ratio (width : radius $\sim$ 1:4) similar to that of HR~4796A and Fomalhaut \citep{kalas_etal05}, and to a model for the cool planetesimal belt shown in Figure~9a of S09.  Given the inferred near-pole-on orientation of the planetary orbits (M08) and the circular symmetry of the resolved far-IR emission (S09), we assume that the HR~8799 debris disk is viewed nearly pole-on. 

For a hypothetical disk inner edge at either 89~AU, in $\sim$3:2 resonance with HR~8799b and coincident with the 90~AU inner edge for the cool planetesimal belt from S09,
the 4$\sigma$ upper limit on the integrated $H$-band disk surface brightness is 15.43~mag (0.69~mJy).
That is, the fraction of dust-scattered light is $f_{\rm scat}<10^{-4}$: a factor of $>$24 fainter than for the HR~4796A disk \citep[$f_{\rm scat}=2.4\times10^{-3}$;][]{schneider_etal99}.
If the HR~8799 debris disk were more extended, similar to that around HD~107146 \citep[width : radius $\sim$ 1:2;][]{ardila_etal04} or as in the preferred model for the cool planetesimal belt in S09 (see their Figure~9b), then $f_{\rm scat} < 2\times10^{-4}$.

Given the strength of the HR~8799 IR excess ($f_{\rm IR} = 2.3\times10^{-4}$),
the ratio of the disk's IR to scattered light is $f_{\rm IR}/f_{\rm scat} > 2.3$.  This compares to $f_{\rm IR}/f_{\rm scat}=1.8$ for HR~4796A \citep[$f_{\rm IR}=4.4\times10^{-3}$;][]{rhee_etal07}.  The higher $f_{\rm IR}/f_{\rm scat}$ ratio for HR~8799 marginally excludes the presence of 
efficient near-IR scatterers between 90 and 240~AU, consistent with: (1) the minimum grain size of several microns set by radiation pressure (in the assumed absence of gas), (2) S09's deduced absence of $<$10~$\micron$ grains in the cool planetesimal belt, and (3) the inferred lack of sub-micron grains in the disk around HR~4796A \citep{schneider_etal99}.

\section{CONCLUSIONS}

We have presented an year-2007 image of all three known HR~8799 planets, including a detection of the inner-most planet HR~8799d 
that precedes its discovery in \citet{marois_etal08b} by one year. Our data exclude the presence of additional $>$3~\Mjup\ outer planets between 68 and 160~AU from the star.
We do not detect scattered light from the debris disk, and place an upper limit of $\sim$18~mag~arcsec$^{-2}$ on its $H$-band surface brightness at 3--6$\arcsec$ (120--240~AU) from the star.
Based on a more robust $H$-band photometric calibration than was possible in \citet{marois_etal08b}, we 
confirm that HR~8799b, c, and d are very under-luminous or redder compared to field brown dwarfs of similar colors or intrinsic luminosities.  We note that this is a distinctive feature of all young substellar objects near the L/T transition, and speculate that it may be due to enhanced dust content and a significant departure from CO/CH$_4$ chemical equilibrium in cool low-surface gravity substellar atmospheres.

\acknowledgments

{\bf Acknowledgments.} 
SM was supported in part by NASA through the {\it Spitzer} Fellowship Program under award 1273192. CM is supported in part through postdoctoral fellowships from the Fonds Qu\'{e}b\'{e}cois de la Recherche sur la Nature et les Technologies.  The data presented herein were obtained at the W.M.\ Keck Observatory, which is operated as a scientific partnership among the California Institute of Technology, the University of California and the National Aeronautics and Space Administration.  The Observatory was made possible by the generous financial support of the W.M.\ Keck Foundation.

\facility{{\it Facilities:} Keck II Telescope}


\end{document}